\pdfoutput=1
\documentclass[10pt]{elsart3-1} 
\usepackage{amssymb}
\usepackage[english,francais]{babel} 
\usepackage[figuresright]{rotating}
\usepackage{amsmath, amsfonts}
\usepackage[most]{tcolorbox} 
\newtheorem{theorem}{Theorem}[section]

\newtheorem{e-proposition}[theorem]{Proposition} 

\newtheorem{e-definition}[theorem]{Definition\rm}

\newtheorem{theoreme}{Th\'eor\`eme}[section]

\newtheorem{proposition}[theoreme]{Proposition}

\newtheorem{definition}[theoreme]{Definition\rm}

\renewcommand{\theequation}{\arabic{equation}}
\numberwithin{equation}{section}
\numberwithin{figure}{section}  
\newcommand \coloneqq {\mathrel{\mathop :}\mathrel{\mkern-1.2mu}=} 
\newcommand{\abs}[1]{\lvert#1\rvert}
\newcommand{\Ric}{\operatorname{Ric}}
\newcommand{\Div}{\operatorname{Div}} 
\newcommand \divt {\operatorname{div}_a}  
\newcommand \Jperp  {\mathbb J^\perp}
\newcommand \Jpar {\mathbb J^\parallel}

\newcommand \Jhatpar {\widehat{\mathbb{J}}^\parallel} 
\newcommand \Pbb {\mathbb P}
\newcommand \Qbb {\mathbb Q}
\newcommand \Jbb {\mathbb J}  
\newcommand \bei 	{\begin{itemize}}
\newcommand \eei 	{\end{itemize}}
\newcommand \del	\partial
\newcommand \auth 	{\textsc}   
\newcommand \Mcal 	{\mathcal M}   
\newcommand \RR 	{\mathbb R}   
\newcommand \Tbb 	{\mathbb T}   
\newcommand \eps 	\epsilon  
\newcommand \be 	{\begin{equation}}
\newcommand \ee 	{\end{equation}} 
\newcommand \bel 	{\be \label}
\newcommand \Kbb 	{\mathbb K} 
\newcommand \dive {\operatorname{div}} 
\AtBeginDocument{\setbox0\hbox{$ $}} 

\journal{the Acad\'emie des sciences}
\begin{document} 

\centerline{}
\begin{frontmatter}
 
\vskip-2.5cm

\selectlanguage{english}
\title{Compensated compactness and corrector stress tensor 
\\
for the Einstein equations in $\Tbb^2$ symmetry}

\selectlanguage{english}

\author[authorlabel1]{Bruno Le Floch}
\ead{bruno@le-floch.fr} 
\and 
\author[authorlabel2]{Philippe G. LeFloch}
\ead{contact@philippelefloch.org}
 
\address[authorlabel1]{Philippe Meyer Institute, Physics Department, \'Ecole Normale Sup\'erieure, PSL Research University, 24 rue Lhomond, F-75231 Paris Cedex 05, France.}

\address[authorlabel2]{Laboratoire Jacques-Louis Lions \& Centre National de la Recherche Scientifique 
\\
Sorbonne Universit\'e, 4 Place Jussieu, 75252 Paris, France.
}

\medskip
\begin{center}
  {\small December 2019}
\end{center}

\begin{abstract}%
\selectlanguage{english}%
We consider the Einstein equations in $\Tbb^2$ symmetry, either for vacuum spacetimes or coupled to the Euler equations for a compressible fluid, 
and we introduce the notion of $\Tbb^2$ areal flows on $\Tbb^3$ with finite total energy. By uncovering a hidden structure of the Einstein equations, we establish a compensated compactness framework and solve the global evolution problem for vacuum spacetimes as well as for self-gravitating compressible fluids. 
We study the stability and instability of such flows and prove that, when the initial data are well-prepared, any family of $\Tbb^2$ areal flows is sequentially compact in a natural topology. In order to handle general initial data we propose a ``relaxed'' notion of $\Tbb^2$ areal flows endowed with a corrector stress tensor (as we call it) which is a bounded measure generated by geometric oscillations and concentrations propagating at the speed of light.  
This generalizes a result for vacuum spacetimes in: {\it Le~Floch B. and LeFloch P.G., Arch.~Rational Mech.~Anal.~233 (2019), 45--86.}
In addition, we determine the global geometry of the corresponding future Cauchy developments and 
we prove that the area of the $\Tbb^2$ orbits generically approaches infinity in the future-expanding regime. In the future-contracting regime, the volume of the $\Tbb^3$~spacelike slices approaches zero and, for generic initial data, the area of the orbits of symmetry approaches zero in Gowdy symmetric matter spacetimes and in $\Tbb^2$ vacuum spacetimes. 
\end{abstract}

%
%
%
%
%
%
%
%
%

\end{frontmatter}

%
   

\vskip-.65cm

\selectlanguage{english} 


 

\section{The global evolution problem for the Einstein equations} 

\paragraph*{\bf Main objective.}

We consider the initial value problem when initial data are prescribed on a spacelike hypersurface and we tackle two major challenges: 
\bei 

\item The presence of {\sl propagating gravitational waves}. These may be impulsive waves in the sense that the spacetime Ricci curvature is a bounded measure (in the presence of matter), while the Weyl curvature may be even less regular. Such waves move at the speed of light and may propagate oscillation and concentration phenomena throughout the spacetime, as pointed out by the authors in~\cite{LeFlochLeFloch-1}.

\item The presence of  {\sl shock waves} ---which generically arise and propagate in a compressible fluid, even when the initial data are regular. This is a classical phenomena in continuum physics due to the nonlinear nature of the Euler equations. Such waves cannot be avoided and a global Cauchy development to the Einstein equations {\sl beyond the formation of shocks} must be sought~\cite{LeFlochRendall-2011}. 

\eei 

\noindent Since we are interested in solutions that may be ``far'' from Minkowski space, it is natural to study this problem first under a symmetry assumption. 
Dealing with spacetimes containing gravitational waves and shock waves requires a new methodology of mathematical analysis. 

Hence, we study here the global dynamics of matter fields evolving under their own gravitational field, and solve the {\sl global evolution problem} for self-gravitating compressible fluids under the assumption of $\Tbb^2$ symmetry, while also providing a significant contribution to the {\sl nonlinear stability and instability} of vacuum spacetimes. This problem was left open after Christodoulou's breakthrough work in the 90's on the global evolution problem for scalar fields in spherical symmetry. We emphasize that propagating gravitational waves are suppressed in spherical symmetry. 

We define here a notion of weak solutions to the Einstein equations and investigate their global geometric properties. 
Our method is based on weak convergence techniques involving energy functionals and compensated compactness properties inspired by Tartar's method \cite{Tartar1,Tartar2}. While bounded variation functions were needed in dealing with spherically symmetric spacetimes, a more involved functional framework is required in $\Tbb^2$ symmetry.  We summarize our results and method in this Note. 

\paragraph*{\bf Einstein-matter spacetimes.}

The global study of matter spacetimes with symmetry was initiated by Rendall~\cite{Rendall-book} and Andreasson~\cite{Andreasson-1999} 
for matter governed by Vlasov's kinetic equation. In this setup, since kinetic matter does not generate shock waves, the existence of global spacetime foliations can be established by arguments similar to those developed for vacuum spacetimes. However, it is significantly more challenging to understand their global geometry, since they can exhibit very different properties in comparison to their vacuum counterparts.  

On the other hand, despite the importance of the problem 
---for instance in astrophysical or cosmological contexts---
of the evolution of self-gravitating {\sl compressible perfect fluids,} the existing (physics only) literature provides  
the construction of special classes of solutions (static fluids) and formal asymptotic analysis only. No rigorous mathematical analysis was available on this problem until recent years. Moreover,  in the mathematical literature only partial results on self-gravitating compressible fluids were available until now, even when attention is restricted to Gowdy symmetry; see \cite{LeFlochRendall-2011}.
Hence, our results are new even in the case of Gowdy symmetry and new also even for vacuum spacetimes. 

\paragraph*{\bf Vacuum spacetimes.}

In the past thirty years, significant progress has been made on the initial value problem for the {\sl vacuum} Einstein equations provided the spacetime metric is {\sl sufficiently regular.} 
The typical symmetry assumption of interest is $\Tbb^2$ symmetry on $\Tbb^3$, that is, 
the existence of two commuting and spacelike Killing fields acting on a spacetime with $\Tbb^3$ spatial topology; 
cf.~Rendall \cite{Rendall-book}.  Such spacetimes admit a global foliation by spacelike hypersurfaces in the areal gauge, that is, such that the area of the orbit of symmetry is a constant on each hypersurface and, furthermore, the corresponding time-function covers the whole range $(0, + \infty)$, describing the evolution originating in a Big Bang up to a forever dispersion toward the future.  The literature on vacuum spacetimes is vast and we refer the reader to \cite{LeFlochSmulevici-2015} for a review and for the theory of future Cauchy developments for $T^2$ symmetric vacuum spacetimes. We emphasize that many very challenging questions still remain open concerning the global geometric behavior of $\Tbb^2$ symmetric vacuum spacetimes: geodesic completeness, curvature blow-up, Penrose conjecture, etc. It is only for the restricted class of Gowdy spacetimes that the global behavior toward the cosmological singularity is now well understood. For instance, we refer to Smulevici \cite{Smulevici-2009} (and the references cited therein) for the strong cosmic censorship in $T^2$ symmetric spacetimes in presence of a cosmological constant.


\section{Weak formulation of Einstein's field equations}

\paragraph*{\bf Einstein's field equations and the Euler equations.}

A spacetime is a $(3+1)$-dimensional Lorentzian manifold $(\Mcal, g)$ satisfying the Einstein equations of general relativity: 
\bel{eq:44}
G = 8 \pi T, 
\ee
which relate, on one hand, the curved geometry of the spacetimes as described by the Einstein tensor $G \coloneqq \Ric - (R/2) g$ and, on the other hand, the matter content of this spacetime represented by the stress-energy tensor~$T$. 
We recall that $\Ric$ is the Ricci curvature tensor associated with the metric $g$ and, by convention, all Greek indices take the values $0, \ldots, 3$.  
A perfect compressible fluid is governed by the stress-energy tensor
\bel{eq:45}  
T = T(\mu,u) = (\mu + p(\mu)) \, g(u, \cdot) \otimes g(u, \cdot) + p(\mu) \, g,  
\ee
where $\mu \geq 0$ denotes the mass-energy density of the fluid and $u$ 
its future-oriented, time-like velocity field normalized to be unit, that is, $g(u,u) = -1$. 
The Einstein equations imply (thanks to the second contracted Bianchi identities associated with the metric $g$) the  Euler equations
\bel{eq:46}
\Div_g T(\mu,u) = 0, 
\ee
in which $\Div_g$ denotes the divergence operator based on the Levi-Civita connection of~$g$. The perfect fluid with stress-energy tensor~\eqref{eq:45} is governed by a general equation of state $p = p(\mu)$, satisfying the hyperbolicity conditions 
\bel{hyperbolic-eos}
p'(\mu) > 0, \qquad 0 < p(\mu) \leq \mu \quad (\text{for all } \mu > 0), 
\qquad p(0) =0. 
\ee
The fluid is described by its mass-energy density function $\mu \geq 0$ and its (unit time-like, future-pointing) velocity vector field $u$. The solutions of \eqref{eq:46}, in general, become discontinuous in finite time even when the intitial data are smooth, so that weak solutions must be sought. 

\paragraph*{\bf Spacetimes with weak regularity.}

Due to the possible formation of shock waves in the fluid variables, the curvature of such spacetimes is defined in the sense of distributions only. To this end, a central issue we investigate~\cite{LeFlochLeFloch-1,LeFlochLeFloch-4} is to find a weak regularity class for the metric and matter variables, under which the global evolution problem for the Einstein equations can be formulated and mathematically solved. 

The first concept of generalized solutions to the Einstein equations was proposed by Christodoulou in a series of papers on spherically symmetric self-gravitating scalar fields. In a 1986 paper and subsequent papers, he investigated solutions to the Einstein-scalar field system in Bondi coordinates, and introduced a class of generalized solutions which are $C^2$ regular {\sl except possibly on the axis} of spherical symmetry. In 1992, Christodoulou introduced a broader class of solutions whose metric coefficients have bounded variation, and established Penrose's weak cosmic censorship conjecture for scalar fields in spherical symmetry. 

In 2007, LeFloch and Mardare \cite{LeFlochMardare-2007} and, more recently, Lott \cite{Lott-2016} gave 
a general definition of the Ricci curvature understood as distributions when the first-order derivatives of the 
metric coefficients are square-integrable. 
We follow \cite{LeFlochRendall-2011} (compressible fluids) and \cite{LeFlochSmulevici-2015} (vacuum spacetimes) and work at the level of weak regularity advocated in~\cite{LeFlochMardare-2007,Lott-2016}. More precisely, our proposal, first made in \cite{LeFlochLeFloch-1} with the more restrictive Gowdy symmetry assumption, is to work within a class of {\sl weak solutions with finite total energy,} 
as we call them. We prove here that the Einstein equations can be solved for arbitrarily ``large'' initial data in such a class of solutions even in presence of a compressible fluid and beyond the formation of shocks. 


\paragraph*{\bf The initial value problem.}

The Einstein equations, together with the Euler equations, can be expressed as a locally-well posed evolution system of partial differential equations of hyperbolic type, provided the pressure $p = p(\mu)$ obeys~\eqref{hyperbolic-eos}.  
For the notion of maximal hyperbolic Cauchy development associated with a given initial data set, we refer to Choquet-Bruhat's textbook~\cite{Choquet-book} and the references therein. For a review of the Cauchy problem in general relativity, see Andersson \cite{Andersson-2004}.
An initial data set for the Einstein-Euler equations \eqref{eq:44}--\eqref{eq:46} consists of a Riemannian manifold $(\Mcal_0, g_0)$
together with a symmetric two-tensor~$k_0$ defined on~$\Mcal_0$, as well as a mass-energy field~$\mu_0$ 
and a vector field~$v_0$ defined on~$\Mcal_0$. These data must satisfy certain constraints, called Einstein's Hamiltonian and momentum constraints \cite{Choquet-book} which we tacitly assume throughout this Note. 
Solving the initial value problem for the Einstein equations (from suitably chosen initial data) consists of finding 
a Lorentzian manifold $(\Mcal, g)$ together with a scalar field $\mu \colon \Mcal \to [0, + \infty)$ and a vector field 
$u$ defined on $\Mcal$, so that the Einstein equations, and therefore the Euler equations, are satisfied in a suitably weak sense while the initial data set is assumed. 


\paragraph*{\bf The $\Tbb^2$ areal foliation.}

The global evolution problem is currently un-tractable by mathematical methods of analysis, and it is natural to study the global problem first under certain assumptions of symmetry. 
Throughout, we assume $\Tbb^2$ symmetry with $\Tbb^3$ spatial topology, and we foliate the spacetime under consideration $(\Mcal, g)$  by spacelike hypersurfaces of constant areal time, denoted by~$t$. Namely, a global time coordinate $t\colon\Mcal \to I \subset \RR \setminus \{0\}$ is introduced that coincides (up to a sign) with the area of $\Tbb^2$-orbits of symmetry.
Here, $I$ denotes an interval that does not contain~$0$, of the form 
$I = [t_0, t_*) \subset (0, + \infty)$ or $I= [t_0, t_*) \subset (-\infty, 0)$.  
We also find it convenient sometimes to state definitions and results on a compact interval $J=[t_0, t_1]$ (not containing the origin). Einstein's constraint equations together with the positive energy condition (enjoyed by our matter model) imply that the gradient $\nabla\abs{t}$ of the area function is a timelike vector field (cf.~\cite{Rendall-book} and the references cited therein). We choose the sign of~$t$ such that $\nabla t$ is future-oriented, thus positive~$t$ and negative~$t$ correspond to future-expanding and future-contracting spacetimes, respectively. 


\paragraph*{\bf Weakly regular, $\Tbb^2$ symmetric initial data sets and Cauchy developments.}

In this context, an initial data set denoted by $(g_0, k_0, \mu_0, v_0)$ consists of data that are defined on the torus $\Tbb^3$, are invariant under a $\Tbb^2$-action and, for simplicity, have constant $\Tbb^2$ area. Likewise, a solution $(g,\mu,u)$ to the Einstein equation is a Lorentzian metric~$g$, a scalar field $\mu\geq 0$ and a unit vector field~$u$, invariant under $\Tbb^2$~symmetry.
Here, $\mu_0 \geq 0$ may vanish and $\mu_0 v_0$ is a spacelike vector tangent to the initial hypersurface, representing the (timelike) matter momentum vector $\mu u$ suitably projected on this hypersurface.  When the density vanishes, the value of the velocity vector~$u$ is irrelevant, and similarly for~$v_0$.
We can view the solution to the Einstein equation as a  $\Tbb^2$ symmetric flow on the torus $\Tbb^3$ consisting of a Riemannian $3$-metric $g(t)$, a two-tensor $k(t)$ (representing the second fundamental form of each spatial slice), 
a lapse function $N(t)= \Omega^{-1}(t)$, a scalar field $\mu(t) \geq 0$, and a vector field $v(t)$ (representing the projection of the physical velocity field). 
In short, $k(t)$~as well as the first-order time and space derivatives of~$g(t)$ are square-integrable on spacelike slices, while the mass-energy density~$\mu(t)$ is integrable and the momentum per particle parallel to the $\Tbb^2$~orbits of symmetry, as well as the lapse function, obey sup-norm bounds. 
At this level of regularity, the Einstein curvature is defined in the sense of distributions, only. 
  
 
\section{Main results} 

\paragraph*{\bf Global geometry of self-gravitating matter.}

For clarity in the presentation, we state a simplified version of our results and refer to \cite{LeFlochLeFloch-4} for more general statements. Our results may apply to fluids governed by pressure laws satisfying the natural hyperbolicity condition~\eqref{hyperbolic-eos} but, for simplicity, we assume that the flow is isothermal, in the sense that its pressure $p(\mu) = k^2 \mu$ depends linearly upon its mass-energy density, where $k \in (0,1)$ represents the speed of sound and does not exceed the speed of light (normalized to unit).  
Since the Euler equations are not reversible in time, 
suitable entropy inequalities are also imposed on the fluid variables, and it is natural to distinguish between two classes of initial data sets, that is, future-expanding and future-contracting spacetimes, respectively. 

\

\begin{theorem}[The global future evolution problem for $\Tbb^2$ areal flows]
\label{thm:1.1}
Consider the initial value problem for the Einstein-Euler equations \eqref{eq:44}--\eqref{eq:46}  
for vacuum spacetimes or isothermal fluids,  
when the initial data set
$(g_0, k_0, \mu_0, v_0)$ is assumed to enjoy $\Tbb^2$ symmetry on $3$-torus topology $\Tbb^3$
and, moreover,  enjoys weak regularity with finite total energy in the sense of Definition~\ref{weakdefinitionT2}
and has a constant area $\abs{t_0}$ (the sign of $t_0\neq 0$ being given below) of either future-contracting or future-expanding type. Suppose that a suitably rescaled component of the fluid momentum parallel to the orbit of symmetry is initially bounded. 
Then, the future Cauchy development of this initial data set is a weak solution $(\Mcal,g,\mu,u)$
of the Einstein-Euler equations with finite total energy.
This solution can be seen as a $\Tbb^2$ areal flow on $\Tbb^3$,
endowed with a foliation by the time function $t\colon\Mcal \simeq I \times \Tbb^3 \to I=[t_0, t_*) \subset \RR$ whose leaves each have topology~$\Tbb^3$ and enjoy $\Tbb^2$~symmetry with orbits of area~$\abs{t}$,
such that: 

\bei 

\item {\bf Future-expanding regime.} One has $t_0>0$ and $t_* = \infty$ and the spacetime foliation extends until 
the volume of the $\Tbb^3$ slices and the area of the $\Tbb^2$ orbits reaches infinity.

\item {\bf Future-contracting regime.} One has $t_0<0$ and the spacetime foliation extends
until an areal time $t_* \in (t_0, 0]$ such that the volume of $\Tbb^3$~slices tends to zero at $t_*$.

\bei 
\item Either $t_*<0$, the length of the $\Tbb^3/\Tbb^2$ quotient reaches zero on the future boundary.  

\item Or $t_* = 0$, that is, the area of the $\Tbb^2$ orbits approaches zero on the future boundary. 

\eei
Furthermore, when the spacetime is vacuum or enjoys Gowdy symmetry, the second case $t_*=0$ generically holds true.   
\eei 
\end{theorem}


\paragraph*{\bf A nonlinear stability and instability theory for the Einstein equations.} 

Our second main result concerns sequences of spacetimes. The statement below is relevant (and new) even for vacuum spacetimes and generalizes the study of vacuum Gowdy spacetimes in Le~Floch and LeFloch \cite{LeFlochLeFloch-1}.  The notions of relaxed areal flow with finite energy and corrector stress tensor are defined precisely below.  

\

\begin{theorem}[The global stability problem for $\Tbb^2$ areal flows]
\label{thm:stability}
Consider a sequence of $\Tbb^2$ symmetric initial data sets $(g_0^n, k_0^n, \mu_0^n, v_0^n)$ (for $n=1, 2, \ldots$) defined on $\Tbb^3$, with the same initial areal time~$t_0$,
and assume their natural norm (cf.~Definition~\ref{weakdefinitionT2}) is uniformly bounded with respect to~$n$, while their volume is uniformly bounded below.
By Theorem~\ref{thm:1.1} the corresponding areal flows $(g^n,\mu^n,u^n)$ are defined on time intervals $[t_0,t_*^n)$.
Up to extracting a subsequence, the maximal time of existence~$t_*^n$ has a limit~$t_*^\infty>t_0$ and the areal flows converge weakly (in a natural norm) to a limit $(g^\infty, \mu^\infty, u^\infty)$ on $[t_0,t_*^\infty)$.

\bei 

\item {\bf Nonlinear stability for well-prepared initial data sets.} 
If the initial data set  $(g_0^n, k_0^n, \mu_0^n, v_0^n)$ converges strongly in the natural norm,  the limit $(g^\infty, \mu^\infty, u^\infty)$ satisfies the Einstein-Euler equations \eqref{eq:44}--\eqref{eq:46}. 

\item {\bf Nonlinear instability for general initial data sets.} 
The limit $(g^\infty, \mu^\infty, u^\infty)$ is a {\sl relaxed areal flow with finite energy} (cf.~Definition~\ref{weakdefinitionT2-deux}), in the sense that it
satisfies an extension of the Einstein-Euler system, obtained by adding to $T_{\alpha\beta}$ a symmetric traceless {\sl corrector stress tensor}~$\tau_{\alpha\beta}$ which is orthogonal to the $\Tbb^2$~orbits of symmetry, is divergence-free and is a bounded measure.
\eei 
\end{theorem} 
  
\

This theorem exhibits a phenomenon in Einstein's field equations (which occurs even in vacuum spacetimes), that is, the appearance of spurious matter terms associated with the gravitational degrees of propagation. 
Christodoulou discovered that singularities in spherically symmetric spacetimes arise at the center of symmetry and do not propagate. On the other hand, in $\Tbb^2$  symmetry, singularities do propagate and require an evolution system of their own. 
Our theorem is a realization of Einstein's intuition that matter could arise as {\sl singularities of the gravitational field.} We refer the reader to the historical discussion in Kiessling and Tahvildar-Zadeh \cite{KiesslingTahvildar} and the references therein.  

 
\section{Methodology} 

\paragraph*{\bf Admissible coordinates.}

In the so-called areal gauge, any $\Tbb^2$ symmetric spacetime metric $g$ on the torus $\Tbb^3$ can be put in the form 
\bel{metric:areal}
  g = \Omega^2 (- dt^2 + a^{-4} dx^2)
  + \abs{t} e^P \bigl( dy + Q\,dz + (G+QH) dx\bigr)^2
  + \abs{t} e^{-P} (dz + H\,dx)^2, 
\ee  
where the metric coefficients $P,Q,\Omega, a, G, H$ with $\Omega,a>0$ are functions of $t \in I$ and $x \in \Tbb^1 \simeq [0,1]$, only, while the remaining  ``transverse'' variables $y,z$ describe $\Tbb^2 \simeq [0,1]^2$.
The metric induced on a $\Tbb^2$-orbit parametrized by $(y,z)$ is
$\abs{t} e^P(dy + Q \, dz)^2 + \abs{t} e^{-P} dz^2
= \abs{t} e^P\,|dy + (Q+ie^{-P}) dz|^2$, which has area~$\abs{t}$ and modular parameter $Q+ie^{-P}$. Our general analysis can be specialized to two cases of interest:
\bei 

\item {\sl Gowdy-symmetric spacetimes} are characterized by the condition $G=H=0$ (everywhere). Geometrically, this is equivalent to the vanishing of the so-called ``twist variables'' ($K_2, K_3$ introduced below).
 
\item {\sl Vacuum $\Tbb^2$ symmetric spacetimes} are defined by taking $\mu = 0$ (everywhere), in which case the Euler equations are automatically satisfied for any arbitrary (and irrelevant) velocity field~$u$. 

\eei 

In order to express the Einstein and Euler equations for the metric~\eqref{metric:areal}, some attention is required to arrive at a tractable system in which the nonlinearities can be analyzed.
We introduce the orthonormal moving frame
\bel{eq:ourframe}
\aligned
& e_0 \coloneqq \Omega^{-1} \del_t, \
&& e_1 \coloneqq \Omega^{-1} a^2 (\del_x - G\del_y - H\del_z), \
&& e_2 \coloneqq | t |^{-1/2} e^{-P/2} \del_y, \
&& e_3 \coloneqq | t |^{-1/2} e^{P/2} (\del_z - Q \del_y).
\endaligned
\ee 
normalized by $g(e_m,e_n)=\eta_{mn}$ with $\eta=\operatorname{diag}(-1,1,1,1)$.
The spin connection components are then essentially ($a/2\Omega$ times) the variables
\be
\Pbb \coloneqq (P_0, P_1), 
\quad
 \Qbb  \coloneqq (Q_0, Q_1), 
\quad
 \Kbb  \coloneqq (K_2, K_3), 
\ee
where
\bel{eq:definevar}
\aligned
P_0 & \coloneqq a^{-1} P_t, 
 \qquad
& Q_0 & \coloneqq a^{-1} e^P Q_t, 
 \qquad
& K_2 & \coloneqq \Omega^{-1} a\, | t |^{1/2} e^{P/2} (G_t+QH_t), 
\\
P_1 & \coloneqq a\, P_x, 
 \qquad
& Q_1 & \coloneqq a\, e^P Q_x, 
 \qquad
& K_3 & \coloneqq \Omega^{-1} a\, | t |^{1/2} e^{-P/2} H_t. 
\endaligned
\ee  
These are our main metric variables. On the other hand, we parametrize the matter content by a momentum vector field $\Jbb=\Omega a^{-1} (2(1+k^2)\mu)^{1/2}u$, where $k$ denotes the sound speed.  We denote its components along the frame~$(e_m)$ by $\Jbb=(\Jperp, \Jpar) = ( J_0, J_1, J_2, J_3)$, so that $\Jperp$ is orthogonal to the orbits of $\Tbb^2$ symmetry while $\Jpar$ is parallel to them. We also define 
$\Jhatpar\coloneqq\abs{a^2\Jbb\cdot\Jbb}^{-q/2}a\Jpar\sim\abs{\Omega^2 \mu}^{(1-q)/2}u^\parallel$, which physically measures the parallel momentum per particle with $q=(1-k^2)/(1+k^2)$.


\paragraph*{\bf The notion of $\Tbb^2$ areal flows.} 
 
We rely here on the notation $\Pbb, \Qbb, \Jperp$ and $\Kbb, \Jhatpar$ above. 

\

\begin{definition}
\label{weakdefinitionT2}
Consider the Einstein-Euler equations with $\Tbb^2$~symmetry on $\Tbb^3$~spacelike slices when the metric is expressed in areal gauge with $t \in I$ describing a compact interval (not containing $0$). Then, a flow of Lebesgue measurable functions 
$t \in I \mapsto (\Pbb, \Qbb, \Jperp, \Kbb, \Jpar)(t)$ together with two real-valued functions $a, \Omega$, defined over a compact spacetime domain $I \times \Tbb^1$ 
is called a {\bf $\Tbb^2$ areal flow on $\Tbb^3$} (or simply a weak solution) provided: 
(1) The following admissibility inequalities hold: 
\bel{eq:admis}
J_0 \geq 0, 
\qquad 
J_0^2 \geq J_1^2 + J_2^2 + J_3^2, 
\qquad
a >0, 
\qquad 
\Omega > 0. 
\ee 
(2) The following integrability and boundedness conditions hold: 
\bel{eq:regul}
\aligned 
\Pbb, \Qbb, \Jperp, a^{-1} & \in L^\infty(I, L^2(\Tbb^1)), 
\qquad
\Kbb, \Jhatpar, a, \log \Omega \in L^\infty(I \times \Tbb^1).
\endaligned 
\ee
(3) The Einstein equations hold in the sense of distributions.  
\end{definition}

\

More generally, for a flow defined on a half-open interval $I=[t_0, t_1)$ we require that the integrability and boundedness conditions hold on every compact subinterval. We validate the above definition by establishing the following statements. 

\

\begin{proposition}
\label{weakdefinitionT2-2-propo}
Under the conditions in Definition~\ref{weakdefinitionT2}, the following properties hold.

\bei

\item Each Einstein equation $G_{mn}=8\pi T_{mn}$, for frame indices $m,n=0,1,2,3$, admits a rescaling by powers of $(\Omega, a)$ that is meaningful in the sense of distributions.  

\item The Euler equations  are meaningful in the sense of distributions and are satisfied as a consequence of the Einstein equations. 

\item The constraint equations enjoy the propagation property: if they hold on a hypersurface of constant time, then they hold within the time interval $I$.  

\eei 
\end{proposition}  


\paragraph*{\bf The notion of relaxed areal flows}   

The previous notion of areal flow is now generalized in order to accommodate {\sl limits of sequences} of solutions to the Einstein equations. All of Einstein's equations hold for these limits, except for the evolution and constraint equations involving the conformal factor~$\Omega$ and additional contributions arises which are {\sl bounded measures,}
 in contrast to the components of the standard matter tensor which are {\sl integrable functions.} 

As we discover, a spurious matter tensor arises which originates in possible oscillations of the geometry itself.  The modified Einstein equations are obtained by adding to the energy-momentum tensor 
a {\bf corrector stress tensor}, as we propose to call it, whose components are bounded measures.
This tensor $\tau=(\tau_{mn})$ is symmetric, traceless, orthogonal to $\Tbb^2$~orbits (in the sense that $\tau_{m2}=\tau_{m3}=0$), satisfies also the positivity condition in 
\bel{eq-posi}
\tau_{00} = \tau_{11}, \quad
\tau_{01} = \tau_{10}, 
\qquad 
|\tau_{01} | \leq \tau_{00},  
\ee
and is {\sl divergence free}
\bel{eq:Twave}
\begin{aligned}
\dive_{a} \big( t \, a \, \tau_{0 \bullet} \big)
= 
 \dive_{a} \big( t \, a \, \tau_{1 \bullet}   \big)
& = 0 .
\end{aligned}
\ee 
Here for a vector field $X = (X_0, X_1)$ we have introduced the notation
$\divt(X_\bullet) \coloneqq - ( a^{-1} X_0 )_t + ( a\, X_1 )_x$,
and we check that all the terms involving the measures~$\tau_{mn}$ are products of continuous functions by measures or are more regular.
We refer to the system consisting of Einstein equations with the matter term augmented with the 
tensor $\tau_{mn}$ as the {\bf relaxed Einstein-Euler system.} 

 \

\begin{definition}
\label{weakdefinitionT2-deux}
Consider the relaxed Einstein-Euler system with $\Tbb^2$~symmetry on $\Tbb^3$~spacelike slices when the metric is expressed in areal gauge, with $t \in I$ describing a compact interval (not containing $0$). Then, a flow of Lebesgue measurable functions 
$t \in I \mapsto (\Pbb, \Qbb, \Jperp, \Kbb, \Jpar)(t)$ 
 together with two real-valued functions $a, \Omega$ and a corrector stress tensor  $\tau_{mn}$, defined over a compact spacetime domain $I \times \Tbb^1$ 
is called a {\bf relaxed $\Tbb^2$ areal flow on $\Tbb^3$}, provided 
(1) The admissibility inequalities \eqref{eq:admis} hold. 
(2) The integrability and boundedness conditions \eqref{eq:regul} hold. 
(3) The corrector stress tensor $\tau$ is a bounded measure and is symmetric, traceless, orthogonal to $\Tbb^2$~orbits, satisfies the positivity condition~\eqref{eq-posi} and is divergence free.
(4) The relaxed version of the Einstein equations holds in the sense of distributions.
\end{definition}

\paragraph*{\bf Weak convergence of null forms.}

By formulating Einstein's field equations as a system coupling nonlinear hyperbolic equations and generalized wave-map equations and relying on a divergence-curl structure that we uncover, we develop arguments of compensated compactness for weak solutions to the Einstein equations as well as for the Euler equations. The relevant structure is found in the following model (presented here in arbitrary spatial dimension ${N \geq 1}$): $\Box \phi_a = Q_a(\del\phi, \del \phi)$ in which $\Box$ is the wave operator in $\RR^{N+1}$ and $\phi = (\phi_a)_{a=1, \ldots, A}$ is the unknown,  while each quadratic term 
$Q_a(\del\phi, \del \phi)$ is a linear combination of the null forms $Q_{0}(\del \phi_a, \del \phi_b) = - \del_t \phi_a \del_t \phi_b + \sum_j \del_j \phi_a \del_j \phi_b$
and $Q_{jk}(\del \phi_a, \del \phi_b) = - \del_j \phi_a \del_k \phi_b + \del_k \phi_a \del_j \phi_b$
(with $1 \leq j,k \leq N$). 
If $\phi^n$ is a sequence of solutions to $\Box \phi^n = Q(\del\phi^n, \del \phi^n)$
and this sequence is uniformly bounded in the $H^1$ norm and 
the sequence of right-hand sides $Q(\del\phi^n, \del \phi^n)$ is $H^{-1}$ compact, 
then its weak limit $\phi^\infty = \lim \phi^n$ is also a weak solution: $\Box \phi^\infty = Q(\del\phi^\infty, \del \phi^\infty)$.  
Namely, under these conditions the div-curl lemma applies and shows that the null forms $Q(\del\phi^n, \del \phi^n)$ are stable under weak convergence, that is, converge to $Q(\del\phi^\infty, \del \phi^\infty)$. 
In our setup, we prove that the sequence $Q(\del\phi^n, \del \phi^n)$ is bounded in the $L^2$ norm and therefore is $ H^{-1}$ compact. We refer to \cite{LeFlochLeFloch-1} for further details.  


\paragraph*{\bf Acknowledgments.}
  
Both authors gratefully acknowledge financial support from the Simons Center for Geometry and Physics, Stony Brook University, at which some of this research was performed. This paper was completed in the Fall 2019 when the second author (PLF) was visiting the Institut Mittag-Leffler for the Semester Program ``General Relativity, Geometry and Analysis: beyond the first 100 years after Einstein''. 


\vspace*{-.15cm}

\bibliographystyle{plain}

\begin{thebibliography}{10}

\bibitem{Andersson-2004}
\auth{Andersson L.,}
The global existence problem in general relativity. The Einstein equations and the large scale behavior of gravitational fields,  
Birkhäuser, Basel, 2004, pp.~71–120.

\bibitem{Andreasson-1999} 
\auth{Andr\'easson H.,}
Global foliations of matter spacetimes with Gowdy symmetry, 
\emph{Commun. Math. Phys.} 206 (1999), 337--366. 

\bibitem{Choquet-book} 
\auth{Choquet-Bruhat Y.,}
{\sl General relativity and the Einstein equations,}
Oxford Math. Monographs, Oxford Univ. Press, 2009.

\bibitem{KiesslingTahvildar}
\auth{M.K.-H. Kiessling and A.S. Tahvildar-Zadeh,}
The Einstein-Infeld-Hoffmann legacy in mathematical relativity I: The classical motion of charged point particles,
\emph{Internat. J. Modern Phys. D} 28 (2019), 1930017. 

\bibitem{LeFlochLeFloch-1} 
\auth{Le Floch B. and LeFloch P.G.,}  
On the global evolution of self-gravitating matter. Nonlinear interactions in Gowdy symmetry, 
\emph{Arch. Rational Mech. Anal.} 233 (2019), 45--86.

\bibitem{LeFlochLeFloch-4} 
\auth{Le Floch B. and LeFloch P.G.,} 
On the global evolution of self-gravitating matter,  
in preparation.  

\bibitem{LeFlochMardare-2007} 
\auth{LeFloch P.G. and Mardare C.,}
Definition and weak stability of spacetimes with distributional curvature, 
\emph{Portugal Math.} 64 (2007), 535--573.

\bibitem{LeFlochRendall-2011} 
\auth{LeFloch P.G. and Rendall A.D.,}
A global foliation of Euler-Gowdy spacetimes with Gowdy-symmetry on $T^3$, 
\emph{Arch. Rational Mech. Anal.} 201 (2011), 841--870.  

\bibitem{LeFlochSmulevici-2015} 
\auth{LeFloch P.G. and Smulevici J.,} 
Weakly regular $T^2$ symmetric spacetimes. The global geometry of future Cauchy developments,  
\emph{J. Eur. Math. Soc. }17 (2015), 1229--1292.

\bibitem{LeFlochStewart-2011} 
\auth{LeFloch P.G. and Stewart J.M.,} 
The characteristic initial value problem for plane-symmetric spacetimes with weak regularity, 
\emph{Class. Quantum Grav.} 28 (2011), 145019--145035. 

\bibitem{Lott-2016} 
\auth{Lott J.,} 
Ricci measure for some singular Riemannian metrics, 
\emph{Math. Annalen} 365 (2016), 449--471. 

\bibitem{Rendall-book} 
\auth{Rendall A.D.,} 
\textsl{Partial differential equations in general relativity,}
Oxford University Press, Oxford, 2008.

\bibitem{Smulevici-2009} 
\auth{Smulevici J.,}
Strong cosmic censorship for $T^2$ symmetric spacetimes with positive cosmological constant and matter, 
\emph{Ann. Henri Poincar\'e} 9 (2009), 1425--1453.

\bibitem{Tartar1}
 \auth{Tartar L.,}
Compensated compactness and applications to partial differential equations, 
in ``Nonlinear analysis and mechanics: Heriot-Watt Symposium'', 
Vol. IV, Res. Notes in Math. 39 (1979), Pitman, Boston, Mass.-London, 136--212.

\bibitem{Tartar2} 
\auth{Tartar L.,}
The compensated compactness method applied to systems of conservation laws,
in ``Systems of Nonlinear Partial Differential Equations'', J.M. Ball ed.,
NATO ASI Series, C. Reidel publishing Col., 1983, 263--285.

\end{thebibliography}

\end{document}